\documentclass[11pt,a4paper]{article}

\usepackage[greek,american]{babel}
\usepackage{indentfirst}
\usepackage[iso-8859-7]{inputenc}
\usepackage{amssymb}
\usepackage[all]{xy}

\newcounter{oftheorem}[section]
\newenvironment{mytheorem}[1]%
{\begin{trivlist}
     \renewcommand{\theoftheorem}{#1~\thesection.\arabic{oftheorem}}
     \refstepcounter{oftheorem}
     \item[\hspace{\labelsep}\bf\thesection.\arabic{oftheorem} #1.]}%
{\end{trivlist}}

\begin{document}
\title{The expressional limits of formal languages in the notion of observation}
\author{Stathis Livadas\\ Department of Mathematics,\\
         University of Patras,
Patras, 26500\\ Greece\\ e-mail:livadas@math.upatras.gr\\tel:+30
6947 302876, fax: +30 2610 997425\\}

\maketitle
\begin{abstract}
In this article I deal with the notion of observation in the most
fundamental sense and its formal representation by means of
languages serving as expressional tools of formal-axiomatical
theories. In doing so, I have taken this notion in two diverse
contexts. In a first context as an epistemic notion linked to an
elaboration of objects of a mathematical theory taken as registered
facts of objective apprehension and then as notion linked to a
process of measurement on the quantum-mechanical level. The second
context in terms of which I deal with the notion of observation is
that of phenomenological constitution basically as it is described
in E. Husserl's texts on phenomenology of temporal consciousness.
Taking that mathematical objects as formal-ontological objects in
phenomenological constitution are based on perceptual objects by
means of categorial intuition, the question is whether and under
what theoretical  assumptions we can, in principle, include quantum
objects in the class of formal-ontological objects and thus inquire
on the limits of their description in the language of a
formal-axiomatical theory. On one hand, I derive an irreducibility
on the level of observables as indecomposable atoms without any
further syntactical content in formal representation and on the
other a transcendence of a continuous substratum self-constituted as
a kind of impredicative objective unity upon which is partly
grounded the definition of an observational frame and the generation
of a predicative universe of discourse.
\end{abstract}

 \vspace{4mm}

 \vspace*{0.4cm}
 \tableofcontents

\vspace*{0.2cm}

 \section{Introduction}
 I generally think that setting the limits of formal languages with respect to
 a notion of observation pretty much of the
 task consists in establishing an underlying all-encompassing theoretical
 background on which to be able to talk about observation either as an
 epistemic notion in the various contexts in which it is encountered
 or as a phenomenological notion linked to {\it a priori} defined acts of an
 `observing' subject. My underlying interpretative scheme will
 be that of a phenomenological constitution of the intentional objects
of experience by a `participating' knowing subject in an
object-like organization of the surrounding
 World-for us which from a certain viewpoint comes close to the version of
 Active Scientific Realism in Quantum Mechanics described
 e.g., in \cite{Karak} and \cite{Karak1}. In my view this interpretation should lead to
 a `representation' of well defined objects in the flux of
 consciousness by means of noetical-noematical constitution which involves a definition of objects as
 reidentifying bearers of predicates across phenomenological time.
 This has to do with a view of objects\footnote{If we regard as phenomenal objects
 those given `primordially in perception' and represented in
consciousness then interpretative
 objects like atoms, electrons, etc. can be also regarded as given by experience
 and thus
 considered as real to the extent that their `reality' is based upon the
 interpretation of sensible signs in an experimental situation
(see: \cite{Heel}, I, p. 5).} in quantum
 mechanical context as general anticipative frameworks in a
 Boolean frame and then in a unifying mathematical - probabilistic tool corresponding to each
 experimental preparation.

 Let me be a little bit more specific here
 about the meaning of the phenomenological terms
 of noetical and noematical described originally
 in E. Husserl's {\it Ideen I} (\cite{Hu1}). A noematical object manifests itself as
 a `giveness' in the flux of a subject's consciousness and it is
 constituted by certain modes of being as such (e.g. with a proper
 to it predicative nest) i.e., as a well defined object immanent to the
 flux which can then be `transformed' to a formal-ontological object and
 consequently a symbolic object of an analytical theory naturally
 including any formal mathematical theory. It can then be said to
 be given apodictically in experience: (1) it can be
 recognized by a perceiver directly as a manifested essence in any
 perceptual judgement (2) it can be predicated as existing according to the
 descriptive norms of a language and (3) it can be verified as such (a
 reidentifying object) in multiple acts more or less at will (\cite{Heel}).

 In
 contrast, an intentional object by hyletical-noetical apprehension ({\it Wahrnehmung})
 can
 be only thought of as an aprioric orientation of intentionality by its sole virtue of being given as such `in person' in
 front of the consciousness inside the open horizon
 of the World-for us.\footnote{The World-for us
 or Life-World in Husserlian terminology can be roughly described to a non-phenomenologist
 as the physical world in its ever receding horizon including in intersubjective sense
 all knowing subjects in a special kind of presence in the World. More on this
 in E. Husserl's {\it The Crisis of European Sciences and Transcendental
 Phenomenology} (\cite{Hu4}).} It is given within a horizon whose
 outer `layer' is the
 boundary between the intentional object and the World-for us; the latter, is
 meant as not being this object or any of its parts and moreover it is
 the `field' of all
next possible noetical apprehensions. These intentional objects as
most fundamental objects of intentionality cannot be reduced to a
lower level of
 apprehension and this is seemingly the reason for which in their
 subsequent temporal constitution as individual noematical objects
 corresponding to `state-of-things'
 they bear no `inner' content at least not one describable by any
 analytical means.

 My interpretational scheme will be also that of a transcendental reduction of
 a phenomenological type when it
 comes to the notion of constitution in-itself as an objectivated,
 homogeneous,
 continuous unity `external' to its immanent objects on which to
 be able to constitute well defined objects of `observation'.\footnote{I note
 here that there exist certain approaches
 claiming a transcendental deduction of Quantum Mechanics ({\bf QM})
 such as M. Bitbol's (\cite{Bitb}) in the sense of a Kantian-type reduction
 of certain particularities of quantum description (e.g. superposition of states, continuity of
 the state vector) to their
 corresponding correlates in internal modes of functioning
 of human consciousness
 or J. Hintikka's ideas on a shift from
 a passive reception of objects to their effective research and instrumental
 shaping on the part of the knowing subject.}
 This continuity in
 the sense of an underlying, impredicative and continuous
 substratum that makes possible to reinstate objects as bearers
 of predicates in kinematical interdependence in each experimental
 context is reduced to a transcendental subjectivity in the self-constitution
 of each one's flux of consciousness
 (\cite{Hu2}, pp. 295-96). I also draw attention onto how
 the unity of a fulfilled time consciousness is reflected in the
language of the
 mathematical theory of Quantum Mechanics in the form of
 classical continuity assumptions in the description of certain
 quantum phenomena or in the form of the state vector representing
 a quantum object
 immediately after measurement
 in a quantum experiment.

 What can be derived from the mathematical
 description of quantum phenomena is the possibility to refer to
 entities as observational and then syntactical individuals in
 disentangled states and further the possibility to describe
 them by means of continuous transformations across time
 (e.g. state vectors or Dirac transformations)  which
 implies, in turn, a continuity of the parameter time. I link
 these two fundamental possibilities in a phenomenologically
 motivated orientation to: (i) The existence of objects of intentionality within
 an outer and
 inner horizon of noetical apprehension - conditioned
 on the existence of a
 relation of intentional character\footnote{This is not to be
understood as a
 relation of some kind of psychological content. By intentional relation,
 which is a phenomenological term, it is
 meant something fundamentally deeper and aprioric. To a non-expert in Phenomenology it can
 be described in one phrase as grounding an aprioric necessity of
 the orientation of a subject's consciousness to the object of its orientation.} between a knowing subject and an
 object - which are then in noematical constitution bearers
 of a fundamental predicative formation
 and (ii) the
 impredicativity of the self-constituting unity of the
 time flux of consciousness leading finally to the transcendence of the pure
 ego in Husserlian terminology. I view these two fundamental irreducibilities
 in subsections
\ref{subsec2} and \ref{subsec3}
 as playing an
 underlying role in the mathematical description of certain quantum
 phenomena such as the Bohm-Aharonov effect and quantum non-separability. An
 approach to
 quantum non-separability motivated to some extent by phenomenological
 concerns has been provided in
 \cite{Karak} and \cite{Karak1}.

 Finally, regarding mathematical objects
 as not
 perceptual objects in literal sense yet founded on
 perception\footnote{In E. Husserl's view, perception by virtue of perceptual acts provides the concrete, immediate and non-reflective basis
 for all our experiences and thus provides also the basis for any intuition of abstract
 objects. For details, see R. Tieszen's \cite{Tiesz1}, pp. 412-15.}
 I look, mainly in section \ref{Sec2}, into how
 the aforementioned irreducibilities are formalized in the language of
 axiomatical mathematical theories, notably in {\it Zermelo-Fraenkel} with
 {\it Axiom of Choice} Theory ({\bf ZFC}), and consequently in the mathematical metatheory of
{\bf QM}. They set, in
 effect, as I shall try to show, the expressional
 limits of a formal language in the notion of observation because
 they stand as irreducibilities of a rather
 phenomenological and certainly of a non-analytical character. By
 this measure, they can be considered as a unifying substratum of
 both quantum-mechanical observation (in the sense of a fundamental
 character `observation') and the logical-axiomatical structure of the
 corresponding formal theory.

 From this aspect, if we adhere to the view that
 mathematical Continuum is primarily based on the intuitive
 Continuum of our real experience and it can be modelized after the
 self-constituted Continuum of the time flux of consciousness
 (which was roughly both L.E.J. Brouwer's and H. Weyl's view of the
 matter, see: \cite{Van Att}), we plausibly expect the transcendental root of
 phenomenological Continuum to be somehow reflected
 in the
 axiomatization of mathematical theories pertaining to a notion of mathematical Continuum. This is what seems to happen with
 the independence from
 the rest of the axioms of {\bf ZF-C} of statements making claims about
 fundamental properties or the cardinality of the Continuum (e.g. {\it Continuum Hypothesis},  the
 {\it Axiom of Choice}, {\it Suslin's Hypothesis}, etc). I simply add as indirectly
 relative to this, G\"{o}del's view that the mathematical essences we
 intuit could not be linguistic conventions in the sense that: ``{\it instead of clarifying the meaning of abstract and non-finitary
 mathematical concepts by explaining them in terms of syntactical
 rules, abstract and non-finitary concepts are used to formulate
 the syntactical rules}" (\cite{Tiesz2}, p. 193).

 \section{Noematical objects as individuals of mathematical theories}\label{Sec2}

It seems to me purposeful to draw attention to what is most
fundamental, in fact what is
 irreducible in the build-up of analytical statements
 of any degree of complexity. If we assume that any
 analytical statement incorporates
 noematical objects\footnote{This kind of noematical objects, e.g. syntactical
 individuals of categorial formulas, numerical symbols, functions of Pure
Analysis, Euclidean or non-Euclidean domains of such functions
etc. were characterized
 by Husserl as referring to `state-of-things' ({\it Sachverhalte}) which are
 intentionalities
 towards an `empty something' ({\it Leeretwas}) (\cite{Hu1}, p.
 33)} in the sense
 of signification objects  ({\it
 Sinnesobjekte}) acting as predicate bearers ({\it Seinssobjekte})
 in noematical constitution together with any doxical modalities
 reflecting consciousness-based states e.g. doubt, certainty,
negation,
 etc., then any attempt at a radical `deconstruction' of the
 analytical structure of any sentence would inevitably reduce to
 statements about individuals and to properties correlated
 with their very nature as individuals. As E. Husserl claimed, those
 sentences are no more of an analytical nature but of a rather
 phenomenological one leading to a kind of `observation'
 of intentional character. This radical reduction which in addition
 to predicate bearers as such reaches their predicative environment as
 well, should fundamentally attach at least the $\in$ predicate as a
noematical correlate to each such predicate bearer in case we talk
about syntactical atoms of analytical - mathematical formulas.

 This type of reduction is meant as the result of
 the elimination of all possible doxical modalities in the
 construction of analytical sentences of any level of complexity as,
 for instance, in general statements expressing
 doubt ($S$ should be $p$), corroboration ($S$ is in fact $p$) or negation
 ($S$ is not $p$) and so on, including also forms uniquely defined by their syntactical
 structure e.g. where one quantifies over elements satisfying
 a particular analytical property $S$ ($\forall p\; S$ or
 $\exists p\; S$). What is left ultimately is a multiplicity of
 individuals (or a collection of such individuals by phenomenological association)
 as intentionally perceived and constituted
 as reidentifying noematical
 objects in a varying predicate environment with
 $\in$ predicate intentionally attached to them as an
 irreducible, non-logical notion
 of order. The predicative nests intentionally attached to
 individuals (in the sense of syntactical atoms) were termed by Husserl as {\it Kernformen} in
 \cite{Hu2} and were supposed to invariably define the essential
 nature of individuals - substrates in the subsequent construction of
 analytical statements of a higher order. From my point of view the $\in$
 predicate is even more fundamental than the equality predicate at
 least in analytical representation for any
notion of equality presupposes a notion of mutual
  inclusion. This is reflected
  in a fundamental way by the
  adoption of {\it Extensionality Axiom}:
  $$\forall X\;\forall Y(\forall u
  (u\in X\longleftrightarrow u\in Y)\longrightarrow X=Y)$$ in the foundations of
  standard set theory as well as in almost any formal theory treating
  sets as definite collections of objects.

  My claim is that this radical reduction to individuals - predicate bearers
  as ultimate phenomenological substrates of analytical statements provides a
  satisfactory interpretational framework for the role of
  urelements of transitive classes under $\in$ predication. As it will become more clear
  in next section, a common foundation
  to both
  formal-ontological objects of categorial
  structures and to objects of quantum `observation' in their
  noematical representation
  in consciousness lies in
  the notion of intentional objects constituted in
  an all-inclusive objectivity  of
  consciousness as reidentifying across time predicate bearers.

  Any intentional object in the possibly lowest level of
  `observation' can, in Husserl's view, be only an
  individual - substrate deprived of any inner structure (even lacking a
  temporal form),
  at least not one
  expressible by any analytical means together with an {\it a
  priori}
  predicative formation by virtue of which it can be perceived
  as a unique noematical object appropriating {{\it eo ipso} a relational property
  with respect to any other such object. It is by all accounts this essential
  characteristic together with the retentional\footnote{The terms
  retention and protention
  are purely phenomenological terms and can be roughly communicated to
  a non-expert in Phenomenology respectively as a kind of immediate conservation in
  memory (retention)
  and immediate expectation of original impression (protention); they are described to
  be of an aprioric and not of psychological character in the
  constitution of the sequence of original impressions in
  the flux of consciousness. For more details, see \cite{Hu3}.}
     } character of the constitution
  process that it is possible to re-identify any
  intentional object of primordial experience
  as the one and same noematical object $x$ under varying
  predicative situations. Any attempt to pass from those individuals perceived
  as a self-donated
  presence in front of the intentionality of consciousness
  to a constituted objectivity of a higher order
  can only entail circularities in description or {\it a priori} terms.
  For instance, in {\it Ideen I} Husserl referred to the multi-ray
  intentionality of the synthetic consciousness which turns by an
  essentially {\it a priori} mode the apprehension of a collection of
  objectivities into an apprehension of
  a single objective whole by what he
  termed a monothetical act whose {\it wesensm\"{a}ssig} (by essence) mode evidently points to a creeping transcendence (\cite{Hu1},
p. 276).

  I'll comment on
  phenomenological transcendence which was
  described {\it in extenso} in terms of the
  self-constitution of time consciousness by Husserl in \cite{Hu3} later in Sec. 4; now
  I call attention to
  irreducible individuals as formal - ontological objects with
  their inherent predicative
  formation and to their implicit role
  in determining properties of transitive classes concerning,
  in particular, the proof of absoluteness of certain categorial
  formulas.\footnote{In rough terms an absolute formula $\varphi$ inside a model $X$ can be
  described as keeping the `mirror' image of itself in any other model $Y$ ordered
  by set inclusion ($X\subset Y$) with respect to the original model $X$.}

  It is very important, for instance, to assure absoluteness of
  certain bounded quantifier set-theoretical formulas in the
  build-up of hierarchies of transitive classes $L_{\alpha}$ in
  G\"{o}del's Constructible Universe L to prove that it serves as a model of
  {\bf ZFC} plus {\bf CH} [{\bf ZF} theory + {\bf AC} ({\it Axiom of Choice}) +
  {\bf CH} ({\it
  Continuum Hypothesis})]. For such formulas the property of absoluteness
  basically is related with the transitivity property of the corresponding
  class; in intuitive terms it has much to do with the invariability of the
  $\in$ - predicative character of the zero-level elements of the original transitive model
  in the recursive definition of classes of any order
  inside it.\footnote{A class $M$ is transitive if for any $x,y,z\in M$
  whenever $x\in y$ and $y\in z$,  $x\in z$. This is equivalent to
the statement
  that whenever $x\in M$
  and $x$ is not a zero-level element (i.e., an urelement under $\in$ predicate)
  then $x\subseteq M$.} For instance, by transitivity property any
  bounded quantifier formula $\varphi$ of the form $$(\exists u\in x)\;
  \psi\;\;\mbox{or}\;\;(\forall u\in x)\;\psi$$ is absolute
  between
  any transitive models $M$ and $N$ whenever formula $\psi$ is.

  The simple proof is based on two assumptions.
  First, that in the inductive definition of absolute formulas
  any atomic formula $\psi$ of the form $i\in j$
  and $i=j$ is absolute and second, that any
  bounded variable $u$ of the formula $\varphi$ is the
  `reflection'
  of a certain invariably the same urelement $u_{i}$ under $\in$ - predication in $M$.\footnote{It is
an immediate
  consequence of the transitivity property of model $M$ that it
  satisfies the Axiom of Extensionality by absoluteness of the
  bounded quantifier formulas $$\forall X\;\forall Y\;[(\forall u\in X)\rightarrow (u\in Y)\wedge
  (\forall u\in Y)\rightarrow (u\in X)]\longrightarrow X=Y]$$ (\cite{Jech}, pp.
  82-83).}

  Both assumptions reduce to admitting the possibility of existence of
  irreducible individuals retaining invariably
  their double nature as individuals-as such and as members
  of the (transitive) class to which they belong. Whether they
  should be objects of stratified categorial formulas in a
  logical - mathematical statement or well defined objects of quantum observation
  expressible as formal - ontological objects in the syntactical norms of a formal - analytical discourse
  changes nothing as to the essence of their individuality
  and their aprioric predication thus leading to a view of them
  as `transformations' of intentional objects of noetical apprehension. The latter case
  (i.e. quantum objects) will be discussed in more detail in the next subsection.

  Whether we may introduce individuals as urelements dropping the Axiom of
  Extensionality as Fraenkel and Mostowski did in constructing
  appropriate models in which the {\it Axiom of Choice} fails\footnote{P. Cohen dismissed those urelements as fictitious objects $x_{i}$
  such that $\forall y\;(\neg y\in x_{i})$ yet $x_{i}\neq x_{j}$
  for $i\neq j$ (\cite{KrauCol}, p. 202). Yet in the sense of noematical individuals
  inside the unity of consciousness that I
  have proposed it makes sense to talk about such objects.} or dismiss them
  altogether reserving this denomination only for null-set ($\emptyset$) yet
  retaining a notion of individuals as
  first-level elements in a cumulative type structure, the
  underlying idea of `indecomposable' individuals preserving invariably
  their syntactical and (in appropriate interpretation) semantical
  content remains fundamentally the same.

  Even viewing urelements of an extended
  Zermelo-Fraenkel universe (ZFU, $\in$) as not identical yet
  indistinguishable elements by the definition of
  $\mathcal{A}$ - indistinguishability inside a
  relational structure $\mathcal{A}=\langle D, \{R_{i}\}_{i\in I} \rangle$ (as proposed by
  Krause and Coelho in \cite{KrauCol}) they can be easily made
  distinguishable by associating to any collection of them an
  ordinal number making thus possible to talk about a collection
   $\sigma_{0},\;\sigma_{1},\;\sigma_{3},....,\;\sigma_{n-1}$ of such
  objects. This is a result of the simple proof that any
  ordinal as a well-ordered structure $\langle A,\;<\;\rangle $ is a rigid
  structure, i.e. the only automorphism in
  this structure is the identity function (\cite{KrauCol}, p. 201).
  In other words, in a rigid structure $\mathcal{A}$ the notion of
  not identical elements and that of $\mathcal{A}$ - distinguishable
  elements coincide. Let us note that the question of the individuality of entities in the
  context of quantum mechanics has provided for much theoretical
  discussion on the nature of quantum objects  as
  they are regarded by some physicists (notably by Scr\"{o}dinger) as
  non-individuals upon which a notion of identity does not make
  sense or by others as bearing a kind of intrinsic individuality by which
  though they might be ``{\small qualitatively the same in all aspects
  representable in quantum mechanical models yet numerically
  distinct}" (\cite{Van Fr}, p. 376).\footnote{In \cite{KrauCol}
  a model-theoretical characterization is proposed of the
   two opposing views on the question of the individuality of quantum
   particles in terms of a trivial and a
    non-trivial rigid expansion
   of a relational structure $\mathcal{A}=\langle D, \{R_{i}\}_{i\in I} \rangle$.
   It is evident by the
   arguments employed in this article that its authors have the view
   that the mathematical structure of
   Quantum Mechanics has a non-trivial rigid expansion (i.e. not one by
   trivially adjoining the ordinal structure) whose
   physical intuition is that quantum entities are somehow
   `intrinsically' distinguishable one from another.}

Individuals as
  purely {\it qua} individuals whether we refer to an axiomatical-mathematical model
  or to a mathematical modeling of quantum mechanics are to
  be thought on a formal level as reflections of irreducible
  intentionalities of consciousness of an {\it a priori} $\;$character
  and they should not be necessarily identified on a formal-ontological level
  with certain elements of standard
  or non-standard theories at least not in
  the absence of the {\it Foundation Axiom} of {\bf ZFC} theory. They can represent any
  entity in the structure of an appropriate formal language as long as it is taken as elementary and not
  further reducible in the sense we have described.\footnote{In \cite{BriRew}, C. Brink and
  I. Rewitzky derive by a proper mathematical modelisation involving
  Priestley duality (something close to Stone duality) that it
  is not essentially different whether we talk either of individuals
  (things), properties, or facts in the world, establishing, in
  effect, an intranslatability between an ontology of individuals
  (nominalism), an ontology of properties (realism) and an
  ontology of facts (factualism).}

  Of course many other relations or properties other than the $\in$$\;$ predicate can be
  predicated to any individual as an object of a formal-ontological discourse but what
  we are most interested in here is to reach the most fundamental,
  the not further reducible level of predication. Evidently, this kind of irreducibility
  connects with a notion of ordinals as a transitive and well-ordered structure
  within a mathematical - axiomatical system.
  Taking into account that by definition an ordinal number is a
  transitive and well-ordered by $\in$ inclusion set it seems
  natural to conjecture that transitive models of {\bf ZFC} should be
  determined by their sets of ordinals. In fact, this was proved
  by P. Vop\v{e}nka and B. Balcar in \cite{Vop} where any transitive
  models $M$, $N$ of {\bf ZFC} are proved to be
  equal ($M=N$) whenever $M$ and $N$ have
  the same sets of
  ordinals with the restriction that {\it AC}$\;$ is satisfied in $M$ ($M\models AC$).

  It is not without importance here that {\bf AC} should be satisfied  at least in
  model $M$ and it gives the motivation to a brief review of this
  independent infinity axiom within the scope of the present work.  The
  intuition of the {\it Axiom of Choice} is that we can, in
  principle,
  apply a criterion of choice at any infinity
  level which would provide us with the possibility to select an
  irreducible individual (think of it as an urelement of a formal theory)
  together with an inherent $\in$
  relation attached to it, among any other potential choices or in
  phenomenological terms among any other possible intentional
  `observations'.

   In a phenomenological approach, I argue that a notion of
   ordering may be automatically induced by any object of intentionality at the
   level of hyletical - noetical apprehension by the sole virtue of the intentional
`property' of the
   object in question to bear an outer horizon,
   i.e. that part of the Life-World that is not the object or parts of the
   object. Evidently, this `property' provides for a complementary
   domain of `observation' for a next potential noetical apprehension
   which by its very enactment provides for a new complementary domain
   and so on. This way a notion of well-ordering may be grounded on
   the noetical level of intentionality with regard to a
   transformation thereafter of an
   hyletical-noetical object to a noematical one
   possibly belonging to
  an aggregation of other such objects in the continuous
  unity of the absolute flux of consciousness. Now, what
  is left after discarding all other details of constitution is
  the possibility to `observe' (and retain) intentionally individuals-as
  such as protentions of intentionality in
  the domain of
  `observation'. By this deconstruction process, a fundamental reduction of the
  {\it Axiom of Choice} as a series of intentional acts of an {\it a priori}
  character inducing in posterior sense a well-ordering among any aggregation of
  formal-ontological objects seems to me plausible.

\section{Can quantum mechanical interpretation
 be related to a phenomenology of constitution?}\label{Sec3}
\subsection{Some remarks on the mathematical language of Quantum
Mechanics}
  In this section I shall argue for the possibility of an interpretation of
  Quantum Mechanics along phenomenological lines especially in
  connection with the notion of the intentional relation
  subject-object and the constitution of
  noematical objects as well defined objects in the self-constituted unity of
  consciousness.  This is an approach that to my knowledge is
  pretty much new
  though there have already been various interpretations beyond the
  mainstream options of realism and instrumentalism, such as M. Bitbol's views in
  \cite{Bitb} motivated by an attempt at a transcendental deduction of
  Quantum Mechanics or the {\it Many-Worlds
   Interpretation} of Quantum Mechanics ({\bf MWI}) taken as  B. De Witt's
   interpretation of H. Everett's `relative state' formulation of {\bf QM}.\footnote{In
   connection with
   this consciousness-related orientation,
   I specifically refer to its `psychological' version
   where the quantum measurement process is roughly reduced to a `splitting' of a
   single consciousness before interaction to several afterwards
   yet retaining by some psychological mode its unity through time.
   See, \cite{Ev1} and \cite{Ev2}.}

 In M. Bitbol's approach I retain, first, his view of an
 original type
 transcendental deduction summed up as providing an internal correlation
 between a unifying mode of appearances of phenomena and certain laws of understanding
 considered as preconditions of experience. This seems, in a quite general form,
 to shift the view towards a field where the mental faculties of a
subject might actively take part in grounding experience as such
and also in shaping up the objects of experience. And, second, I
think we should keep in mind the meaning of his constraint of
contextualization which corresponds a Boolean subframe to each
experimental preparation linked to a unified mathematical tool of
probabilistic prediction irrespective of the context associated to
the measurement that follows the preparation (\cite{Bitb}, p. 11).
In such a case,
 based on the notion of a reidentifying object across time (which
 in my view presupposes
 the existence of an otherwise
 irreducible intentional relation subject/object) we can
 ascribe to each experimental preparation a unified, predictive
 (non-Kolmogorovian) tool whose valuations are associated with a Boolean
 framework irrespective of the context associated to the measurement that
 follows the preparation.

This seems interesting to the extent that: a) it presupposes an
object-like organization of phenomena to be described in the
language of a Boolean observer `attached' to each experimental
situation and b) it implies a unifying mathematical tool bridging,
in effect, the contextual frames of the preparation of an
experiment and its measurement.

The former condition can be understood as leading to the following
assumptions. The indirect introduction of a participating
`observer' who has a particular mode of `observation', then a
particular mode of constituting his observations and reproducing
them in a predicable object-like universe and a particular
language to describe them as well defined objects. In the case of
an experimental preparation and measurement this is inherently
linked with an observer's  capacity to transform his
intentionalities in terms of noetical-hyletical apprehensions of
the real world to noematical objects in terms of reidentifying
objects across time and thus well defined bearers of predicates in
an ordered context. Then he would be able to talk, in principle,
about the ontological nature of these constituted objects
irrespectively of whether they are considered as objects under
predication of a formal or of a common natural language. For they
have become objects of a formal ontology which in reverse order
reduce by phenomenological intentionality to individuals- as such
bearers of an outer horizon in hyletical - noetical apprehension.

In view of the aforesaid I propose an interpretation (in terms of
a noetical - noematical constitution) of the presence of a knowing
subject that performs quantum measurements via a measuring
apparatus in the following fashion: The measured property produces
a macroscopic effect on the instrument (e.g. a pointer reading or
a track in a bubble chamber) which is a material sign. This can be
considered as having a double reality; its material one as a
pointer sign or a bubble track and an intentional reality proper
to it as a sign susceptible to be constituted at a next stage as a
formal-ontological object. A sign regardless of its particular
material content has the mode of being a sign-as such and in being
so it can be thought of as an intentional object of noetical
apprehension by virtue of being merely a so-called
`state-of-things' ({\it Sachverhalt}), in other words an `empty
something'\footnote{By the transformation of originally given
intentional objects as `state-of-things' to formal-ontological
objects it is possible for a physical interrelation to be
formalized mathematically with numerical (or mathematical) symbols
corresponding to observable signs furnished by the intermediary of
a measuring instrument.}; it should then be apriorically directed
to a knowing subject performing the experiment by means of a
measuring device. In any case, it may be assumed that the signs of
a measuring apparatus are symbols of certain physical properties
(natural symbols) insofar as they are uniquely determined by the
interaction with a quantum entity in terms of which they
`translate' the hidden state of the quantum entity into uniquely
determined sensible signs (\cite{Heel}, p. 174-175). Evidently,
these signs which are part of the `physicalistic language' of the
measuring apparatus can be considered as intentional objects for
the performing subject who can then turn them to linguistic
symbols; the latter is conditioned on his capacity to constitute
them as well defined noematical objects in his flux of
consciousness. As symbols of a linguistic statement they are not
just material reality signs but they are parts of a predicative
environment by being symbols as-such corresponding to
`state-of-things' which are moreover bearers of two important
properties: 1) They do not determine a unique linguistic
statement. By being symbols-as such they can be arguments of
equivalent logical-mathematical formulas inasmuch as they are
abstractions of unique material signs and 2)  They are devoid of
any inner analytical content as they are linguistic symbols
abstracting in each case a unique and irreducible intentional
object, e.g. the sign-as such of the bubble track of a particle.

At the stage a knowing subject will be able to represent
noematical objects as such and such well defined objects of
discourse in such or other internal noematical mode he should by
necessity have constituted them already in a kind of synthetic
unity to be able to talk about them together at once; and this
unity should undeniably be a temporal unity. But arguably this
reduces to a constitution of internal time in the form of a
continuous unity of time consciousness which has almost nothing to
do with external (or scientific) time e.g. the time frame of
classical or relativistic theory.

In fact, both quantization conditions and the continuous wave-like
propagation of phenomena, stemming from the formalism of Quantum
Mechanics with appropriate boundary conditions, are due to the
intrinsic property of quantum objects to be `embeddable' as
outcomes of sufficiently reproducible experiments to a unified
meta-contextual frame of probabilistic description. This
possibility  involves, on a phenomenological level, at once both a
relationship subject-object of an intentional character and the
noematical constitution of intentional objects as well defined
objects in the unity of consciousness; in a quantum experiment
this is translatable to an embedding of reproducible
`observations' in a meta-contextual Boolean subframe.\footnote{My
intentional/constitutional approach is not directly connected with
the general stance one might have on the question of the nature of
quantum particles as there is no unconditioned description of
them. Anyway I note that in the
      light of EPR critique and the antinomies of entanglement
      states Schr{\"o}dinger attempted a reinterpretation
      of the epistemological questions of quantum theory and a
      reexamination of the question of the individuality of
      particles.
 In fact, it was his
      attempt to interpret the formal results of the
      Bose-Einstein statistics which implicated an indiscernibility of monoatomic
      gas molecules that led him to abandon the particle
      interpretation and adopt the undulatory view in terms of
      which the gas as a physical system should be considered as a
      system of stationary waves in which molecules are just states of
      excitation energy deprived in this way of individuality.
      Later he gave a physical interpretation in
      electromagnetic sense
to the wave function
      $\psi$ as solution of his general equation but that was again
      of a wave configuration
      consisting in the superposition of all  kinematically possible
      point-configurations of the system each one intervening by
      its special ``weight" in the physically interpreted formula $\int \psi
      \psi^{\ast}$ (current density).  Moreover his
      wave image proved more satisfactory in representing atomic
      transitions by an energy exchange between different
      vibrations rather than by a quantum leap between states in
      which case one cannot possibly describe the transition in time and
      space.}
 In addition and independently of the context that follows the preparation of
 a quantum experiment there should be some intrinsic way by which
 quantum objects as intentional ones become reidentifying objects of a constituting
consciousness invariably over (internal) temporal unity.
Eliminating then all time-related modes of noematical constitution
(e.g., simultaneity, succession, casual relationship) there should
be a temporal substratum of the predicable universe of Quantum
Mechanics  whose temporality should be something radically
different than the ordinary objective time of a classical
macroscopic system. This may lead to argue that the unifying,
meta-contextual time of the predictive tool should be the
objectivated `reflection' of the absolute flux of time
consciousness whose objectivation cannot be described but as a
sort of `mirror' reflection of its ever {\it in-act} self.

I close this subsection by referring to a well-known quantum
effect where the derivation of quantum conditions by classical
continuity assumptions as constraints provides a clue to the
necessity of assumption of an impredicative temporal substratum to
which I just referred above.

 In the case of the potential well, for instance,
 the discrete eigenvalues of the energy operator is  the
 formal result of continuity assumptions about the wave function $\psi$ on
 the boundaries of the potential well and of constraints put on
 the wave function out of the classically permitted region,
 in the `observational limit' to infinity.\footnote{Based on the
 continuity of the wave function of a free particle at the boundaries
 $x= \pm \frac{a}{2}$
 of a potential well $V$  we get the equations $\psi(\frac{a}{2})= \psi(-\frac{a}{2})$ and
  $\psi'(\frac{a}{2})= \psi'(-\frac{a}{2})$. For $x< -\frac{a}{2}\;$ or $x> \frac{a}{2}$, in the  limit at
infinity it
  must hold that ${\mathop{\mathrm{lim}}\limits_{x\longrightarrow \pm \infty}}\;\Psi(x)=0$
  for the wave function $\Psi$ of a free particle of energy $E$ (with $E< V$) which in this
  region of the plane takes the form of a descending exponential function
  $\psi(x)=C \exp(k_{1}x) + D \exp(-k_{1}x)$ (see \cite{Mess}
  or any other basic {\bf QM} textbook).}

   Both
   constraints underlie an observational capacity linked at least
   indirectly to a subjective notion of temporal continuum. In that sense,
   the condition of continuity at $x= \pm \frac{a}{2}$ of the
   wave function and its derivative implies the underlying existence of a
   continuous substratum of internal time providing a continuous domain
   for the particle state function whereas normalization
   condition ${\mathop{\mathrm{lim}}\limits_{x\longrightarrow \pm \infty}}\;\Psi(x)=0$
   which is in accord with the classical intuition that the formal representation of
   the physical state of a free particle should approach zero at infinite
   limit, is a classical limit equation presupposing a continuous
   space-time substructure.
    It seems that a continuous substratum of time-consciousness which
    is also asserted (the primordial intuition of mathematics) in a non-standard notion of mathematical
   Continuum\footnote{For a brief survey of L.E.J. Brouwer's and H. Weyl's modelization
   of mathematical Continuum after the phenomenological Continuum I refer
   to M. Van Atten's {\it et al} work in \cite{Van Att}.}
   should be implicitly assumed in setting classical
   limit equations ${\mathop{\mathrm{lim}}\limits_{x\longrightarrow \pm \infty}}\;\Psi(x)=0$ and equations of continuity of the wave
   function $\Psi$ at the boundaries  $x= \pm \frac{a}{2}$.

   To sum up, discrete eigenvalues of bound states of a quantum
   system are in mathematical formulation partly an indirect consequence of classical
   limit and
   continuity assumptions against an underlying substratum of impredicative
   continuum of propagation. An objectivated,
   time-fulfilled Continuum ({\it erf\"{u}llte Kontinuum}) must be also
   presupposed in the phase of second (or field) quantization in
   the relativistic version of Quantum Mechanics where single
   particle wave functions of classical version are transformed
   into quantum field operators on quantum states defined on any
   space-time point. This is basically implemented by
   an extension of Langrangian formalism to field equations.

   To demonstrate the underlying presence of this kind of impredicative continuous
   spatio-temporal substratum in the mathematical theory of quantum mechanics I refer, in the following
   subsection, to the well-known Bohm-Aharonov effect.
   \subsection{The case of the Bohm-Aharonov
   effect}\label{subsec2}
   The main purpose of my reference to the Bohm-Aharonov effect
   is to show that certain irregularities on the
   observational - physical level are reduced in mathematical tool to special topological
   properties of the relevant configuration space.  In the specific effect
   the irregularity has to do with the presence of a solenoid
   causing a shift in the interference pattern of a double
   slit in the notable absence of an external magnetic field. Moreover,
   the physical effect observed which is the change in phase
   difference in the electron interference pattern
   $\Delta\delta=\frac{e}{\hbar}\int \mbox{curl}A\;dS$ depends only
   on curlA\footnote{The magnitude A is the vector potential which in classical
   physics, as it is well-known, is
   linked to the magnetic induction B by the formula $B=\mbox{curl}A$.} so that it
   could be deduced that an electron is influenced by
   fields which are only non-zero in regions inaccessible to it.
   In formal terms, this amounts to a non-locality of the integral
   $\oint Adr$. In short we could say that the Bohm-Aharonov effect owes
   to the non-trivial topology of the vacuum (in this particular
   case the space outside the solenoid) and the fact that
   electrodynamics is a gauge theory (\cite{Ryd}, p. 101).

   Being a bit more specific, without intending to enter into
   the details of the experimental context, the existence of the Bohm-Aharonov
   effect is essentially translatable to a topological situation where the
   configuration space of the null
   field is a plane with a hole in it, that is the
   non-simply connected circle $S^{1}$. In further mathematical
   elaboration, this generates a many-valued gauge function $x$
   mapping the group space $S^{1}$ onto the configuration space of
   the experiment $S^{1}\times R$ such that not all such $x$
   are deformable to a constant gauge function ($x=\mbox{const})$. In
   that case, it would produce $A_{\mu}=0$ and no Bohm-Aharonov
   effect (\cite{Ryd}, p.
   105). In mathematical formalism the function $x$ such that
$A=\nabla x$ turns out to be
   a many-valued function and this becomes possible since the space in which it is defined is not simply connected. That
   is, the group space of the gauge group of electromagnetism $U(1)$ is the
   non-simply connected circle $S^{1}$ where, roughly speaking,
   a non-simply connected space is one in which not all curves may be continuously shrunk to a point.

   If $x$
   were single-valued, then $B=\mbox{curl}A=\mbox{curl}\nabla x\equiv 0$ everywhere, so there would
   be no magnetic flux at all and consequently no physical effect taking into account
   that $\Delta\delta=\frac{e}{\hbar}\Phi$.

   In view of our previous discussion, we note in this specific quantum
   mechanical experiment a `transformation' of the irregular
   observational characteristics of the quantum phenomenon into peculiarities in the
   topological texture of a
   spatiotemporal continuous substratum; in the particular
   case the peculiarity lies in the property of non-simply connected
  of the configuration space of the experimental
   context. But, in generating topological
   properties leading to certain discontinuities in configuration
   space one must assume, prior to the assumption of discontinuity gaps
   in topological structure, the constancy of an underlying spatiotemporal continuum
   across time
   which can in
   turn reduce to the constancy of a fulfilled time-consciousness self-constituted as a
   continuous unity `bridging', in effect, the context of an experimental preparation
   with that of measurement.

   \subsection{Interpreting quantum
   non-separability}\label{subsec3}
   In quantum mechanical theory quantum non-separability arises first
   as a result of
   the principle of superposition of states and second
   from the impossibility to provide, given a compound
   system S and its corresponding Hilbert space H, a decomposition
   of it
   into a tensor product $H=H_{1}\otimes
   H_{2}\otimes ...\otimes H_{N}$ of the subsystem spaces $H_{i}$
   such that an observable A of S can be expressed in the canonical
   form $A=A_{1}\otimes A_{2}\otimes ... \otimes A_{N}$ of
   suitable observables of the subsystems $S_{i}$. Formally this
   is a result of the particular feature of the tensor product
   that it is not a restriction of the topological product
   $H=H_{1}\times
   H_{2}\times ...\times H_{N}$ but includes it as a proper subset. Given that in
   quantum mechanical theory there
   are no reasonable criteria that would guarantee the existence (and uniqueness) of such
   a tensor product decomposition of the whole system the question
   is how we could possibly derive it and on what terms on an
   operational level.\footnote{A prototype of an EPR - correlated system experimentally
confirmed is the compound
   system S of spin-singlet pairs. It consists of a pair $(S_{1},
   S_{2})$ of spin $\frac{1}{2}$ particles in the singlet state $$
   W= \frac{1}{\sqrt{2}}\;\{\mid \psi_{+}> \otimes \mid\phi_{-}> -
   \mid \psi_{-}> \otimes \mid \phi_{+}>\},$$ where $\{\mid
   \psi_{\pm}>\}$ and  $\{\mid
   \phi_{\pm}>\}$ are orthonormal bases of the two dimensional
   Hilbert spaces $H_{1}$ and $H_{2}$ associated with $S_{1}$ and
   $S_{2}$ respectively. In such a situation, it is theoretically
   predicted and experimentally confirmed that the spin components
   of $S_{1}$ and $S_{2}$ have always opposite spin
   orientations.}

    V. Karakostas, for instance, discusses in \cite{Karak} and \cite{Karak1} the question of
   non-separability from the point of view of Active Scientific Realism as presupposing the feasibility of the kinematical
   independence between a component subsystem of interest and an
   appropriate measuring system including its environment; it presupposes, in
   general, the separation between the observer and
   the observed. Taking the physical world as an unbroken whole we have to
   separate it, to perform a breakdown of the entanglement of
   subsystems. In what is called a Heisenberg cut,
   we have to decompose the compound entangled system into interacting but
   disentangled components that is, into measured objects on the one
   hand and measuring systems (uncorrelated observers in a broad
   sense) on the other with no (or insignificantly so) holistic correlations among
   them. By means
   of the Heisenberg cut can be generated well-defined separate objects in their
   contextual environments described in terms of a process of projecting the holistic
   non-Boolean domain of entangled quantum correlations into a
   meta-contextual Boolean frame that breaks the
   wholeness of nature by means of an effective participancy in
   the physical world of a knowing/intentional subject (\cite{Karak}, pp. 300,
   303). In fact, the notion of an effective participancy of a
   knowing/intentional subject in the physical world seems to
   imply the Aristotelian idea of {\it potentia} since, on a quantum level, for any
   effective observer  inside the Life-World there should be
   two categories of entities, those posterior to
   his knowing/intentional acts which as already pointed out he has some inherent mode to recognize
   as well-defined objects and those prior to his purely intentional acts which
   should by necessity be for him mere potentialities; in that sense, ``{\small
a quantum
   object exists, independently of any operational procedures,
   only in the sense of `potentiality', namely, as being
   characterized by a set of potentially possible values for its
   various physical quantities that are actualized when the object
   is interacting with its environment or a pertinent experimental
   context}" (\cite{Karak1}, p. 290).

   In view of description of the relation
   between a knowing subject and an object of his intentionality (in terms
   of noetical-noematical constitution), offered mainly in subsection
   3.1, we may argue that there exists a certain convergence of the
   interpretational content of phenomenological analysis with the
   positions of Active Scientific Realism\footnote{In a certain sense this
   approach is related to H. Everett's `Relative State Interpretation' of
   Quantum Theory e.g. by means of a decoupling of world components
   $\psi^{(R)}$, $\psi^{(L)}$
   of a certain superposition state $\psi(t)= e^{iHt} \phi (\varphi_{R}\pm
   \varphi_{L})$
   corresponding to a localization of consciousness not only in space and time but
   also in certain Hilbert space components (see example: \cite{Zeh}, pp. 73-74).} inasmuch as:

   The implementation of the Heisenberg cut
   can be taken in a fundamental sense as presupposing a notion of
   co-existence and also an idea of separation in a domain of
   intentional `observation'
   between a consciousness intentionally directed to its object
   and the object in-itself as a direct and unambiguous presence
   in front of the intentionality of consciousness. By applying
   his intentionality a knowing subject creates a particular
   context to inquire e.g. on the `hidden status' of an entangled
   quantum state in the following two stages: 1) on the noetical level by apprehending a
   sensible sign (of the measuring apparatus) as such in the sense
   that it could not be otherwise but apprehend it as a sign
   distinguishable from any other possible sign in the
   protention of his intentionality (cf: \cite{Hu5}, p. 8);  at this stage he has already
   lost his claim on an access to the inner reality of the
   entangled state for he noetically apprehends what he apprehends in the
   `physicalistic language' of the apparatus and 2) thereafter he
   constitutes it as a noematical object immanent to his consciousness in the
   modes already described.

   Moreover, I feel that the introduction of the effective
   presence of a knowing/intentional subject in the Life-World
   puts on a close footing these two interpretations with
   respect to the aristotelian notion of potentiality as they
   seem to somehow weaken the vaguely metaphysical character of
   this principle exactly by the introduction of an
   intentional/constituting subject as part of the Life-world. So, from a
   phenomenological point of view, a World in which
   pre-predicative structures (i.e. intentionality structures) linked to the
   presence of an intentional/constituting subject
   determine by `anticipation' actual predicated instances may be defined
   as a domain of real possibility anterior to actuality. This
   seems partly to eliminate the vague ontological status - not to
   say purely metaphysical - of Aristotelian potentialities
   for it substitutes for the notion of a first {\it entelechy},
   reached necessarily by regression ad infinitum of all classes
   of potentialities, the notion of at least one constituting
   subject in a pre-phenomenological World. Connected to the
   view above, is the assertion that the inherently probabilistic nature
   of Quantum Mechanics  may be interpreted as due to the
   irretrievable loss of information caused by the cut of a quantum
   non-separable whole in the measuring process. This means that in view of
   the reduction
   to the intentionalities of a subject performing a quantum
   experiment certain potentialities of a quantum whole
   are realized whereas others are not on the level of noetical apprehension
   and this is what can,
   in principle, be asserted for any particular
   contextual experimental frame.

   From this aspect, the principle of actualization put up
   by R. Omn$\acute{e}$s as an additional external rule not emerging
   from the internal structure of  {\bf QM} to postulate the
   passing from phenomena to facts and used ``{\small merely as a licence to
   use consistent logic to reason from present brute experience}" (\cite{Far},
   p. 1335) leads at least indirectly to an effective participancy of a
knowing/intentional subject.

   In phenomenological approach as the
   knowing/intentional subject acting as a constituting factor
   transposes the pre-phenomenological `unity' of
   fundamental experience to the a-thematic, impredicative
   unity of its self-constituted flux of consciousness (\cite{Hu2}, pp. 283, 295)
   it looks as if this underlying impredicative spatio-temporality should bear its
   `imprint' on the interpretation of the wholeness of a quantum
   non-separable state standing as an undissectable whole and a limit to a complete
   scientific cognizance of physical reality. A prime reason for this limitation
   may lie in the fact that we lack any possible way
   to `go deeper' than intentionality and consequently the analytical means
   to fully describe
   the inner time of an entangled state before or exactly at the stage of
   noetical apprehension; more
   generally it seems that we lack the means to unconditionally approach the
   temporality of the pre-phenomenological World (the World before the
   phenomenological
   reduction of the constituting
   {\it Ich}) which in Husserl's writings
   is presupposed as the constant synthetic unity of every possible experience
   and also the common
   denominator in terms of substance of all beings in the World.
   This phenomenological
   `incompleteness' might somehow account for the inherent impossibility to
   provide a complete description of the World as a whole by means
   of a formal and logically consistent theory that would also include its universe (including the
   knowing/intentional subject) as its own object. Just as any
   language of an axiomatical system of mathematics capable of
   expressing at least elementary arithmetic cannot but eventually
   engender antinomies or paradoxes (cf. with G\"{o}del's First Incompleteness Theorem) especially in relation to
   self-referential descriptions. In quantum world we could claim
   that this runs parallel to the example of von Neumann's account of
   quantum measurement that leads to an infinite regressional sequence of observing observers
   (\cite{Karak}, p. 306).

   In this connection, I refer to M.L.
   dalla Chiara's view of the measurement problem of quantum
   mechanics as a characteristic question of the semantical
   closure of a theory, in other words as to ``{\small what extent a
   consistent theory (in this case {\bf QM})  can be closed with respect to
   the objects and the concepts which are described and expressed
   in its metatheory}". According to dalla Chiara, quantum mechanical theory
   as a consistent theory satisfying some
standard formal requirements,
   turns out to be the subject of
   some limitations due to purely logical reasons concerning
   its capacity to completely describe and express certain physical
   objects and concepts.

   Nevertheless, even if a contradiction produced in
   the metatheory of {\bf QM} can be overcome on the purely logical
   grounds (linked to similar limitative results on the consistency of axiomatical
   systems in set theory) that
   ``{\small any apparatus which realizes the reduction
   of the wave function is necessarily only a metatheoretical
   object}" (\cite{Chiar}, p. 338) the question, in my view, remains open of
   providing a consistent and complete metatheoretical description
    as to what `happens' in physical state terms in-between the
    experimental preparation of a compound
   system such as $s\otimes \mathcal{Q}$ and the time of measurement corresponding to
   the collapse of the wave function  (where $s$ is a physical
   state at time $t$ and $\mathcal{Q}$ a measuring
   apparatus identified with a Boolean-minded observer assigning truth values
   to non-Boolean quantum substructures). The jump of truth values
   in the process of measurement which is formally the result of
   the absence of an isomorphism between Boolean and non-Boolean
   structures - assuming that a quantum object, considered as an objective
   existence, is the
   non-distributive lattice of its properties - forces for a Boolean observer the
   need of the existence
   of an objective time in which he must `move' (\cite{Grib}, p.
   2396).

   This question is also linked with J. von Neumann's Projection Postulate (or `The Reduction
   of the Wave Function' postulate) as it implicitly establishes the necessity of a
   self-constituting time flux by assigning to the mathematical
   translation $\tau((s)(t))$ of a physical state $s(t)$ at time $t$ the same
   eigenvector $\psi_{\kappa}$
   as for the measured quantity
   $Q_{i}$ of the state $s(t)$ at time $t_{1}$ soon after the
   measurement.\footnote{By this postulate we get as a result of measurement the interval
   $r_{\kappa}\pm \epsilon_{Q_{i}}$, where $r_{\kappa}$ is an eigenvalue
   of the mathematical form of $Q_{i}$ and $\psi_{\kappa}$ its
   corresponding eigenvector (\cite{Chiar}, p. 344).} As a matter of fact, even if we
   assume Von Neumann's Projection
   Postulate or Van Fraassen's modal interpretation of Quantum Mechanics
   as `external'
   metatheoretical conditions in a purely logical way we
   cannot be led by any linguistic means to a complete description of the `physical
   change' that takes place
   during the measurement process in the compound system
   `system + apparatus'.
   This raises
   again the question of a self-constituting time flux of
   consciousness and the constitution of objects in it as noematical
   correlates of hyletical-noetical moments of an outward directed
   intentionality.

   Closing the section I turn again to the fundamental
   irreducibilities which in my view shape in an essential way
   our observational frame in an intersubjective world of an
   unbounded horizon of events: On the one hand intentionalities of an
   {\it a priori} character directed on the lowest level of primordial experience
   to individuals - as such transposed then  with their
   noematical correlates as immanences of the flux of each subject's
   consciousness. On the other, the intuition of continuous unity
   as a substratum divested of any quality on which to constitute
   and deliver a meaning to well-defined noematical objects described
   deeply enough as an
   impredicative self-constituting unity of the flux of consciousness leading
   ultimately to a
   transcendental ego of consciousness (see: \cite{Hu3},  pp. 97-99).

   Saying it in
   more intuitive terms, as much it is impossible to reduce the
   mental process by which we may abstract from an original impression in immediate
   awareness evidently distinct from any other
   in temporal flow to anything more fundamental in
   noetical apprehension it is equally impossible to capture what is
   constituted as the unity of a whole in consciousness by means
   of the former activity.

\section{Observation in the language of
formal systems.
 Where is the irreducibility and where the transcendence?}
 As the main purpose of this article was to discuss the limits of
 formal languages with respect to the notion of observation I
 naturally sought to reach the most fundamental level of
 observation beyond the limits of the common intuition of this notion. In doing
 so, I took also into account the claim, which was E.
 Husserl's belief, that mathematical objects are special cases of
 perceptual objects leaving aside any counter-arguments which are
 nevertheless of a rather artificial nature, e.g. whether the
 mathematical object $\{\emptyset\}$ should be also considered  a
 perceptual object. My theoretical standpoint, linking fundamental observation to a
 phenomenology of constitution put under the same perspective the mental process
 of formation of mathematical
 objects as syntactical atoms corresponding to `state-of-things'
 in a formal-ontological environment and the
 process of constitution of quantum entities as well defined noematical objects
 in consciousness based on their former intentional apprehension
 in the physical world.\footnote{My approach in this paper is not meant to
 be a transcendental deduction of Quantum
Mechanics
   by means of Phenomenology for, notwithstanding my claim to
   providing
   some clues on a possible phenomenological interpretation of  {\bf
   QM} on the level of `observation' and in metatheory, there are certain
   constants (the Planck constant, for instance) or symmetry principles of {\bf
   QM} which are still doubtful whether they are of a purely
   empirical objective character or possibly susceptible of a phenomenological interpretation.
   Nevertheless, for the Planck constant, for example, there is a view of it as a not purely
   extrinsic datum but as arising from the generic situation of
   mankind which in my opinion leads indirectly to a notion of
   intersubjective constitution in the Life-World.} It is obvious that in
such an approach we should regard mathematics as divested of
 any platonist content and in a certain sense devoid of the
 conveniences of Cantorian-type actual infinity. In this
 connection, mathematical theories of an alternative nonstandard
 character especially those which incorporate e.g. a notion of
 natural infinity as an open-ended shift of classes of
 hereditarily finite `observations' (for instance, Alternative Set
 Theory and
 Hyperfinite Set Theory) seem more adapted to my view of
 mathematical activity as a special kind of abstraction in an
 intersubjective and interactive field of events of a `local'
 but ever receding horizon.

 This way, talking of an object as an individual-as such it is in
 a fundamental and essential way the same whether it is a
 syntactical atom of a stratified mathematical formula corresponding
 to a unique `state-of-things'
 or whether it is a
 quantum entity of a certain non-separable state apprehended by
 a disentangled interaction and
 transformed as a reidentifying-noematical
 object by a knowing/intentional subject using a measuring apparatus
 as an extension of his consciousness.
As long as they can be apprehended as distinct to any other
possible intentional
 apprehensions in the process of constitution they can both be
 classified as individuals and if in addition they can be
 constituted
 as bearers of an otherwise undefined sense of `order' to any
 other such apprehension they can be classified as
 individual-substrates bearers by essence of an appropriate to them predicative
 environment ({\it Kernform}). Is there a way to penetrate more deeply,
 to open and `read' the inner content of those individuals,
 in a word, to reach a deeper level of apprehension? The answer
 seems to be negative and in addition not in the sense of a contingent state of
affairs but of a generic state of affairs. The main indication is
our own intuition in the direct givenness of the intentional
objects of our experience and this is why the intentionality of
 primordial experience is described exactly as intentionality to
 intentional objects, i.e. to individuals-as such. I would call it a most
 fundamental irreducibility in relation to human perception though
 it is at the same time rather `friendly' and easy to co-operate with our
other mental faculties. It is thanks to this fundamental intuition
that we can comprehend and handle almost anything from sequences
of  natural numbers to the capacity to shape images of various
distinct particle trajectories in a bubble chamber.

 On the contrary, there is another irreducibility which though it is our most common intuition
 it proves most difficult to comprehend let alone describe by formal
 first-order  linguistic means. This is the intuition of
 Continuum which includes everything from the common intuition of
 our existence as a continuous state of events, to the intuition
 of a curve on a piece of paper as a continuous set of black
 points,
 to the intuition of subatomic events as taking place against a
 time-fulfilled continuous background. What is it that makes possible this
 coherent unity in constitution which is reflected as a formalized continuous substratum
 of the mathematical metatheory of `observations' irrespective of whether they
 refer to a quantum mechanical context or to a context of common
 physical intuition?

 Husserl made a clear distinction
 between phenomenological time, the homogenous form of all living
 experiences in the flux of consciousness and the objective or
 scientific time. By necessity every real
 experience is a durating experience which is a fact extracted by pure
 intuition of its enactment and it is constituted by
 a certain {\it a priori} mode as an infinitely fulfilled
 continuum of durations. Going
 deeper into the ontology of phenomenological Continuum
 Husserl encountered grave difficulties in comprehending it as he
 himself professed in {\it Ding und Raum} and he left it even in his
 later writings as a rather obscure notion leading him to a
 transcendental pure ego of consciousness. This ego as an absolute and
 impredicative subjectivity is only accessible by its objectivation
 as a `mirror image' of itself and it is only in this way possible to reflect
 on the Continuum as an objective whole and also on the notion  of an unbounded infinity in a Kantian sense
 (\cite{Hu1}, p. 331). Essentially Husserl
 eradicated the transcendence of the world of the purest idealist
 doctrine only to introduce it backdoor by means of a personified pure
 ego.

 The matter, in last count, is not so much whether one
 should in principle accept a transcendence of intuitive and
 consequently of mathematical Continuum by means
 of a phenomenological reduction or by some other interpretational
 scheme as the hard fact that any attempt to describe Continuum by
 the first-order linguistic means of a formal axiomatical system
 inevitably leads to circularities in definitions or entails some form of {\it ad hoc}
 axiomatization. This is presumably reflected in the independence
 of actual infinity principles such as {\it Continuum Hypothesis} and of the {\it
 Axiom of Choice} from the other axioms of the
 Zermelo-Fraenkel Set Theory and to some extent in the {\it ad
 hoc}
 extension or prolongation principles axiomatizing the embedding of standard
 structures into saturated nonstandard domains (see: \cite{Lev}).

 There is an
 ongoing theoretical discussion on the possibility of
 a non-analytical character of the
 {\it Continuum Hypothesis}
 question in the foundation of mathematics and on this account we refer
 to S. Feferman's thesis in \cite{Fef} that ``{\small the
{\it Continuum
 Hypothesis} is an inherently vague problem that no new axiom will
 settle in a convincingly definite way}".

 It
  seems worthwhile to close by mentioning a recent
  neuroscientific approach to the constitution of unitary and
  coherent experiences out of independent bits of data in which
  the processes of predication and identity between different
  occurrences of variables (viewed as individuals) are
  reflections of the same underlying conceptual process in the
  brain. In \cite{Piet} this process is described as diagrammatic
  and iconic rather than symbolic, in a certain sense as a neuronal
  continuous connection between differently localized predicates
  which involves also {\it Gestaltpsychological} notions
  substituting traditional symbolic operations.

  But invoking some sort of continuous or
  topological representation in the context of neuronal processes described by
  first order
  formalism one might get
  trapped once more in the inherent impredicativity of the notion of
  Continuum. It is like swinging to the other end of the pendulum
  as we mark the following irreducibilities its extreme points; the
  intentionality to individuals-as such on the one hand and the
  intuition of the impredicative Continuum on the other.

 The question of the nature of Continuum as well as the epistemic content of the
 aprioric principle of intentionality
 might be the object of a yet deeper research linking such diverse
 fields as mathematics, logic, quantum theory including its offspring
 quantum gravity, cognition theory and neurophysiology of the
 brain.

 Or they might in principle be elusive and beyond any reach on the grounds of
 the circular question: On what terms can mind capture the mind?

\end{document}